\documentclass[twocolumn,showpacs,aps,floatfix]{revtex4}

\usepackage{amsmath2000}
\usepackage{graphicx}% Include figure files
\usepackage{dcolumn}% Align table columns on decimal point
\usepackage{bm}% bold math

\begin{document}
\preprint{ND Atomic Theory 2002-3}
\title{Mixed configuration-interaction and many-body perturbation theory calculations of energies and oscillator strengths of J=1 odd states of neon}
\author{I. M. Savukov}
 \email{isavukov@nd.edu}
 \homepage{http://www.nd.edu/~isavukov}
\author{W. R. Johnson}
 \email{johnson@nd.edu}
\homepage{http://www.nd.edu/~johnson}
\author{H. G. Berry}
 \email{Berry.20@nd.edu}
 %\homepage{http://www.nd.edu/~johnson}
\affiliation{Department of Physics, 225 Nieuwland Science Hall\\
University of Notre Dame, Notre Dame, IN 46566}

\date{\today}
\begin{abstract}
\textit{Ab-initio} theory is developed for 
energies of $J=1$ particle-hole states of neutral neon and for oscillator
strengths of transitions from such states to the $J=0$ ground state. 
Hole energies of low-$Z$ neonlike ions are evaluated. 
\end{abstract}

\pacs{31.10.+z, 31.30.Jv, 32.70.Cs, 32.80.-t}
\maketitle

\section{Introduction}

A combined configuration-interaction (CI) many-body-perturbation-theory
(MBPT) method, applied previously to divalent atoms~\cite{isav5}, is extended to
particle-hole states of closed-shell atoms. 
After derivation of CI+MBPT expressions for
particle-hole states, we will apply the theory to calculations of energies
and electric-dipole transition probabilities for neon.

For neon, many accurate
measurements of transition rates are available, providing important tests of
theory. Reciprocally, the theory might help resolve existing
discrepancies among oscillator strengths ($f$-values) for transitions from the
ground state to several excited states, for which experiments disagree.
There is also a certain deficiency in existing \textit{ab-initio} theories in neon,
for which discrepancies among many measurements and theoretical calculations
are unsettled. For example, the only other elaborate \textit{ab-initio}
calculations (\citet{neon:98}) give an oscillator
strength for the $\left[ 2p_{3/2}^{-1}3s_{1/2}\right] _{1}$ neon state larger than
most experimental values by more than two standard deviations. Extensive
calculations performed by \citet{neon:93} for many
transition rates along the neon isoelectronic sequence use a general
configuration-interaction code (CIV3)~\cite{neon:75}. The calculations utilize
parametric adjustments with measured fine structures, 
but do not completely agree with experiments in neon and
have an accuracy similar to other semiempirical calculations of~\citet{neon:98a}. 
% [Seaton98].
However, the two calculations disagree with each other for several
transitions. We hope that our calculations may help
to understand better the theoretical problems in neon and provide guidance
for the analysis of experimental data.

Some possible applications of the present CI+MBPT method include the study
of neonlike ions, Ne I -- Si V, S VII, Ar IX, Ca XI, and Fe XVII that have
astrophysical interest and have been included in the Opacity
Project (\citet{opac}). 
The transition data in neon and other noble gases are also used in
plasma physics, and in studying discharges that find many industrial
applications in lamps and gas lasers. The methods presented here might be
also used for improving the accuracy of MBPT or for extending CI+MBPT to
more complicated open-shell atoms.

The principal theoretical difficulty arises from the sensitivity of transition
amplitudes to the interaction between closely spaced fine-structure
components. Although it is possible to obtain energies which are 
reasonably precise on an
absolute scale using coupled-cluster methods (\citet{kaldor}), 
accurate fine-structure splittings seem very difficult to
obtain without semiempirical adjustments. This is why semiempirical
approaches, which have fine-structure intervals carefully adjusted, are more
successful in neon than are \textit{ab-initio} calculations. However, as we
will demonstrate in this paper, CI calculations corrected with MBPT are also
capable of accurately predicting fine-structure splittings and,
consequently, transition amplitudes. In this paper, we will
demonstrate the excellent precision of CI plus second-order MBPT. 
Third-order corrections, 
for which numerical codes already exist~\cite{neon:92},
can also be included, providing even further improvement in accuracy.

In the following section, we use the effective Hamiltonian formalism
and particle-hole single-double coupled equations to derive expressions for
the second-order Hamiltonian matrix of the CI+MBPT method. In the final
expressions, we present a quite accurate new MBPT that can predict
energies of hole states and can describe appropriately the interactions in
particle-hole atoms. The accuracy of hole energies obtained with the new
MBPT will be illustrated for neon and  low-$Z$ neon-like ions. Our CI+MBPT
energies and $f$-values for many states of neon are
tabulated. Their agreement with experiment and other theories are shown.

\section{CI+MBPT method}

The accuracy of the Rayleigh-Schr\"{o}dinger variant of second-order MBPT
given in \cite{neon:01} is insufficient for purpose, so that more accurate
single-double equations must be used. The formulas for the correlation
operator and a system of coupled equations for the correlation coefficients
are given in \cite{neon:95}; we follow the notation of \cite{neon:95} in the
the paragraphs below. Under certain conditions, those equations
can be further simplified and rewritten in the following form: 
\begin{eqnarray}
\left( \varepsilon _{b}-\varepsilon _{\alpha }\right) \chi _{b}^{\alpha }
&=&R_{b}^{\alpha } \nonumber \\
\left( \varepsilon _{b}+\varepsilon _{c}-\varepsilon _{\alpha }-\varepsilon
_{\beta }-\widetilde{g}_{bcbc}\right) \chi _{bc}^{\alpha \beta }
&=&R_{bc}^{\alpha \beta }-\widetilde{g}_{bcbc}\chi _{bc}^{\alpha \beta } \nonumber \\
\left( \varepsilon _{v}-\varepsilon _{r}\right) \chi _{v}^{r} &=&R_{v}^{r} \nonumber \\
\left( \varepsilon _{a}-\varepsilon _{v}\right) \chi _{v}^{a} &=&R_{v}^{a} \nonumber \\
\left( \varepsilon _{v}+\varepsilon _{b}-\varepsilon _{r}-\varepsilon
_{s}\right) \chi _{vb}^{rs} &=&R_{vb}^{rs} \nonumber \\
\left( \varepsilon _{v}+\varepsilon _{b}-\varepsilon _{a}-\varepsilon
_{s}\right) \chi _{vb}^{as} &=&R_{vb}^{as}\text{.}
\end{eqnarray}
In the second equation of this set, the term $\widetilde{g}_{bcbc}\chi
_{bc}^{\alpha \beta }$ is subtracted from both sides of this equation to
make the right-hand side small. Since large
random-phase approximation (RPA) corrections in the particle-hole CI+MBPT 
are treated by CI, the
quantities $W_{va^{\prime }}^{v^{\prime }a}$ entering this set of equations
on the right-hand side in Ref.~\cite{neon:95} %[Avg95]
are small and have been neglected here. The concern might be raised for the
correlation coefficients $\chi _{vb}^{as}$ and $\chi _{v}^{r}$, which
generally would have small factors $\left( \varepsilon _{v}+\varepsilon
_{b}-\varepsilon _{a}-\varepsilon _{s}\right) $ or $\left( \varepsilon
_{v}-\varepsilon _{r}\right) $ in front. However, for the large CI model
space, energies of the core-virtual orbitals $bs$ are well separated from
the energies of the valence-hole orbitals $av$. The quantities $R$ in zero
approximation can be set to: 
\begin{eqnarray}
R_{j}^{i} &=&\Delta _{ij} \nonumber \\
R_{bc}^{ij}-\tilde{g}_{bcbc}\chi _{bc}^{ij} &=&g_{ijbc} \nonumber \\
R_{vb}^{is} &=&g_{isvb}
\end{eqnarray}
to obtain the first-order effective Hamiltonian, 
\begin{equation}
H_{v^{\prime }a^{\prime },va}^\text{eff}=(\varepsilon _{v}-\varepsilon
_{a})\delta _{v^{\prime }v}\delta _{a^{\prime }a}+H_{v^{\prime }aa^{\prime
}v}^{(1)}
\end{equation}
and the correlation coefficients $\chi $. Here we define the first-order
correction $H_{v^{\prime }aa^{\prime }v}^{(1)}=\Delta _{v^{\prime }v}\delta
_{a^{\prime }a}+\widetilde{g}_{v^{\prime }aa^{\prime }v}$ to the effective
Hamiltonian. For faster convergence of CI and for subtraction of the
dominant monopole contributions in RPA diagrams, a $V^{(N-1)}$ Hartree-Fock
(HF) model potential for which $\Delta _{nm}=\widetilde{g}_{nama}$, $\Delta
_{na}=\Delta _{an}=\Delta _{ab}=0$ is introduced.

Further improvement of accuracy can be achieved through iterations. After
one iteration we obtain the second-order contribution to the effective
Hamiltonian, 
\begin{equation}
H_{v^{\prime }aa^{\prime }v}^{(2)}=\delta R_{v}^{v^{\prime }}\delta
_{a^{\prime }a}+\delta R_{a}^{a^{\prime }}\delta _{v^{\prime }v}+\delta 
\widetilde{R}_{va}^{av^{\prime }}\text{,}
\end{equation}
where 
\begin{eqnarray}
\delta R_{v}^{v^{\prime }}&=&\sum_{s\notin CI}\frac{\Delta _{v^{\prime
}s}\Delta _{sv}}{\varepsilon _{v}-\varepsilon _{s}}\nonumber \\
&& -\sum_{scd}\frac{g_{cdvs}%
\widetilde{g}_{v^{\prime }scd}}{\varepsilon _{c}+\varepsilon
_{d}-\varepsilon _{v^{\prime }}-\varepsilon _{s}-\widetilde{g}_{cdcd}} \nonumber \\
&&+\sum_{stc}\frac{g_{v^{\prime }cst}\widetilde{g}_{stvc}}{\varepsilon
_{v}+\varepsilon _{c}-\varepsilon _{s}-\varepsilon _{t}}\text{,} \\
\delta R_{a}^{a^{\prime }}&=&-\sum_{scd}\frac{g_{cda^{\prime }s}\widetilde{g}%
_{ascd}}{\varepsilon _{c}+\varepsilon _{d}-\varepsilon _{a}-\varepsilon _{s}-%
\widetilde{g}_{cdcd}} \nonumber \\
&&+\sum_{scd}\frac{g_{cda^{\prime }s}\widetilde{g}_{ascd}%
}{\varepsilon _{a^{\prime }}+\varepsilon _{c}-\varepsilon _{s}-\varepsilon
_{t}-\widetilde{g}_{a^{\prime }ca^{\prime }c}}\text{,} \label{eq6}\\
\delta \widetilde{R}_{va}^{av^{\prime }} &=&\sum_{tu}\frac{g_{av^{\prime }tu}%
\widetilde{g}_{tuva^{\prime }}}{\varepsilon _{v}+\varepsilon _{a^{\prime
}}-\varepsilon _{t}-\varepsilon _{u}} \nonumber \\
&& +\sum_{cd}\frac{\widetilde{g}%
_{cdva^{\prime }}g_{av^{\prime }cd}}{\varepsilon _{c}+\varepsilon
_{d}-\varepsilon _{a}-\varepsilon _{v^{\prime }}-\widetilde{g}_{cdcd}} 
\notag \\
&&+\sum_{t\notin CI}\frac{\widetilde{g}_{av^{\prime }ta^{\prime }}\Delta
_{tv}}{\varepsilon _{v}-\varepsilon _{t}} +\sum_{dt\notin CI}\frac{%
\widetilde{g}_{dv^{\prime }ta^{\prime }}\widetilde{g}_{atvd}}{\varepsilon
_{v}+\varepsilon _{d}-\varepsilon _{a}-\varepsilon _{t}} \nonumber \\
&&-\sum_{dt}\frac{\widetilde{g}_{dv^{\prime }tv}\widetilde{g}_{ata^{\prime
}d}}{\varepsilon _{a^{\prime }}+\varepsilon _{d}-\varepsilon
_{a}-\varepsilon _{t}-\widetilde{g}_{a^{\prime }da^{\prime }d}} \nonumber \\
&&+\sum_{t\notin CI}\frac{\Delta _{v^{\prime }t}\widetilde{g}_{taa^{\prime }v}%
}{\varepsilon _{v}+\varepsilon _{a^{\prime }}-\varepsilon _{t}-\varepsilon
_{a}}  \notag \\
&&+ \sum_{dt}\frac{\widetilde{g}_{datv}\widetilde{g}_{v^{\prime }ta^{\prime
}d}}{\varepsilon _{a^{\prime }}+\varepsilon _{d}-\varepsilon _{v^{\prime
}}-\varepsilon _{t}-\widetilde{g}_{a^{\prime }da^{\prime }d}} \nonumber \\
&&-\sum_{dt}\frac{%
\widetilde{g}_{data^{\prime }}\widetilde{g}_{v^{\prime }tvd}}{\varepsilon
_{v}+\varepsilon _{d}-\varepsilon _{v^{\prime }}-\varepsilon _{t}-\widetilde{%
g}_{adad}}. 
\end{eqnarray}
Note that in the last equation we have extended the single-double method.
The last term entering $\delta \widetilde{R}_{va}^{av^{\prime }}$ in the
single-double formalism would normally not contain $\widetilde{g}_{adad}$ in
the denominator. However, if we do not modify this denominator, we find that
in the third-order MBPT, large terms proportional to $\widetilde{g}_{adad}$
will appear leading to a decrease in accuracy. A physical reason for
modifying the denominator of this term is that the process described by this
term contains two holes in the intermediate states with large interaction
energy. This interaction should be treated nonperturbatively, for example,
by inclusion of $\widetilde{g}_{adad}$ into the denominator as we have done
on the basis of the single-double equations in other terms. Finally, this
term is almost equal to the seventh term (they are complex conjugates and
their Goldstone diagrams are related by a reflection through a horizontal
axis), and for convenience they are set equal in numerical calculations. The
angular reduction for $\delta \widetilde{R}_{va}^{av^{\prime }}$ can be
easily obtained using the second-order particle-hole formulas given in Ref.~\cite{neon:01}.

\section{A solution of the hole-energy problem}
\subsection{Breit corrections}
Apart from Coulomb correlation corrections, the Breit magnetic interaction
is also important in neon and the isoelectronic ions. The breakdown of
various Coulomb and relativistic contributions to the energy of $3s$ states
of neon are given in Ref.~\cite{neon:95}. Breit corrections cancel, but for
higher excited states they may not. Hence, to improve the accuracy of 
fine-structure splittings, we include the Hartree-Fock hole Breit correction 
$B_{aa}^\text{(HF)}$ in our calculations, 
\begin{equation}
B_{aa}^\text{(HF)}=\sum_{c}\widetilde{b}_{acca}\text{.}
\end{equation}
We have checked that the first-order corrections $B^{(1)}$
 to the energies of $J=2$ and $J=1$
states given in Table I of Ref.~\cite{neon:95} agree with our $B_{aa}^\text{%
(HF)}$ contributions, 0.00062 and 0.00090 a.u., for $2p_{3/2}$ and $%
2p_{1/2}$ states, respectively. 
We omit the small
frequency-dependent Breit, quantum-electrodynamic, reduced-mass, and
mass-polarization corrections. 
Small as they are, those corrections are further reduced 
after subtraction for the fine-structure intervals. 
More careful treatment of
relativistic corrections is needed in calculations of high-$Z$ neon-like ions.

\subsection{Calculations of hole energies for neonlike ions}

Since we propose a new variant of the MBPT expansion, we would like first to
demonstrate that this expansion is convergent for hole states. The
theoretical hole energies shown in Table~\ref{Table:hole} have been obtained 
in the $V^{(N)}$ HF potential using Eq.~(\ref{eq6}) for $\delta R_{a}^{a}$ 
to calculate second-order corrections. The extra term in the denominator is important and
is necessary for convergence of the perturbation expansion. Experimental
hole energies in the National Institute of Standards and Technology (NIST)
database Ref.~\cite{nist:01} are found as the limit energies for the neon
isoelectronic sequence. For neutral neon only one limit, the p$_{3/2}$ energy is
given in NIST~\cite{nist:01}. The 2p$_{1/2}$-2p$_{3/2}$ splitting 780.4269(36) cm$^{-1}$
has been measured in Ref.~\cite{neon:85}, and using this value we find the
experimental p$_{1/2}$ energy. Table~\ref{Table:hole} demonstrates the good agreement of
our theoretical p$_{3/2}$, p$_{1/2}$ energies as well as the same fine
structure interval for neon-like ions. Our fine structure interval, whose
correctness is crucial for transition amplitude calculations, differs from
experiment just by about 10 cm$^{-1}$. Note that the HF value 187175 cm$%
^{-1}$ for the 2p$_{3/2}$ state is 8.5\% higher than the experimental value
173930 cm$^{-1}$, and, after adding correlation corrections, we obtain
improvement by a factor of ten. For the fine structure, the HF value 1001cm$%
^{-1}$ disagrees even more, by 28\%. If we use Rayleigh-Schr\"{o}dinger
perturbation theory, the corrections are twice as large as our results, and
the agreement with experiment does not improve.

\begin{table}[tbp]
\caption{A comparison of theoretical and experimental hole energies and the
 2p$_{3/2}$-2p$_{1/2}$ fine-structure intervals for neon and neon-like ions.
All energies are in cm$^{-1}$}
\label{Table:hole}
\begin{center}
\begin{tabular}{lrrrrr}
\hline\hline
& Ne & Na$^{+}$ & Mg$^{+2}$ & Al$^{+3}$ & Si$^{+4}$ \\ \hline
2p$_{3/2}$ Th. & 172434 & 380443 & 645951 & 967531 & 1344344 \\ 
2p$_{3/2}$ Exp. & 173930 & 381390 & 646402 & 967804 & 1345070 \\ 
Difference & 1496 & 947 & 451 & 273 & 726 \\ \hline
2p$_{1/2}$ Th. & 173218 & 381816 & 648196 & 970997 & 1349449 \\ 
2p$_{1/2}$ Exp. & 174\thinspace 710 & 382756 & 648631 & 971246 & 
1350160 \\ 
Difference & $\allowbreak $1492 & 940 & 435 & 249 & 711 \\ \hline
2p$_{3/2}$-2p$_{1/2}$, Th. & 784 & 1373 & 2245 & 3466 & 5090 \\ 
2p$_{3/2}$-2p$_{1/2}$, Exp. & 780 & 1366 & 2229 & 3442 & 5105 \\ 
Difference & -4 & -7 & -16 & -24 & -15 \\ \hline
\end{tabular}
\end{center}
\end{table}

\section{Neon energies and oscillator strengths of J=1 odd states}

To test the accuracy of the CI+MBPT method, we first calculated energies of
several lowest odd J=1 neon states, Table~\ref{tab:neonen}. The number of
configurations in CI was chosen to be 52. The order of eigenstates obtained
in CI+MBPT is the same as the order of the experimental levels. We
abbreviate long NIST designations since the levels are uniquely specified by
energy or by order.

\begin{table}[tbp]
\caption{A comparison with experiment of CI+MBPT energies referenced to the
ground state and given in atomic units. An almost constant shift is
subtracted in the fifth column to demonstrate excellent agreement for
relative positions of levels}
\label{tab:neonen}
\begin{center}
\begin{tabular}{ccccc}
\hline\hline
Level & Experiment & CI+MBPT & $\Delta$ & $\Delta$ - 0.0069 \\ \hline
$p_{3/2}^{-1}3s$ & 0.6126 & 0.6048 & 0.0078 & 0.0009 \\ 
$p_{1/2}^{-1}3s$ & 0.6192 & 0.6116 & 0.0076 & 0.0007 \\ 
$p_{3/2}^{-1}4s$ & 0.7235 & 0.7166 & 0.0070 & 0.0001 \\ 
$p_{1/2}^{-1}4s$ & 0.7269 & 0.7200 & 0.0069 & 0.0000 \\ 
$p_{3/2}^{-1}3d$ & 0.7360 & 0.7289 & 0.0070 & 0.0001 \\ 
$p_{3/2}^{-1}3d$ & 0.7365 & 0.7294 & 0.0071 & 0.0002 \\ 
$p_{1/2}^{-1}3d$ & 0.7401 & 0.7330 & 0.0071 & 0.0002 \\ 
$p_{3/2}^{-1}5s$ & 0.7560 & 0.7491 & 0.0069 & 0.0000 \\ 
$p_{1/2}^{-1}5s$ & 0.7593 & 0.7525 & 0.0069 & 0.0000 \\ \hline
\end{tabular}
\end{center}
\end{table}
The pure \textit{ab-initio} energies differ from experimental energies by
0.0069 a.u., but after subtraction of the systematic shift (which does
not make much difference in transition calculations),
the agreement is at the level of 0.0001 a.u.  for almost all states. 
Therefore, we consider the
accuracy of CI+MBPT adequate for correct prediction of level mixing and
oscillator strengths. For the 3s states, agreement with experiment for the
fine structure interval is much better than that obtained by~\citet{neon:95}, 
0.0002 versus 0.0012 a.u.; a possible
explanation for this could be that single-double equations miss important
corrections which we included by modifying the denominators. In Ref.~\cite{neon:95}, 
however, the systematic shift is small. 

Finally, we present our CI+MBPT oscillator strengths in neon. 
After diagonalization of
the second-order effective Hamiltonian, we obtain wave functions in the
form of expansion coefficients in the CI space and use them to calculate
oscillator strengths. Size-consistent formulas for dipole matrix elements
for transitions decaying into the ground state are provided in Ref.~\cite{neon:98}, 
where the absorption oscillator strength $f$ is also defined. We
give in this table \textit{ab-initio} values of the oscillator strengths $f$. 
The dominant part of the RPA corrections is included at the level of CI.
Small normalization corrections are omitted.

\begin{table}[tbp]
\caption{Our CI+MBPT oscillator strengths for the ground to excited state
transitions in neon compared with average experimental values (3rd and 4th columns) 
and those
obtained with the best semiempirical theories~\cite{neon:98a,neon:83,neon:93}}
\label{tab:neontr}
\begin{center}
\begin{tabular}{lcccccc}
\hline\hline
Levels & CI+MBPT & $\sigma$-avr & mean & Ref.~\cite{neon:98a} & 
Ref.~\cite{neon:83} & Ref.~\cite{neon:93} \\ \hline
$p_{3/2}^{-1}3s$ & 0.0102 & 0.0099 & 0.0107 & 0.0126 & 0.0106 & 0.0123 \\ 
$p_{1/2}^{-1}3s$ & 0.1459 & 0.1549 & 0.1487 & 0.1680 & 0.1410 & 0.1607 \\ 
$p_{3/2}^{-1}4s$ & 0.0131 & 0.0122 & 0.123 & 0.0152 & 0.0124 & - \\ 
$p_{1/2}^{-1}4s$ & 0.0181 & 0.0170 & 0.016 & 0.0193 & 0.0160 & - \\ 
$p_{3/2}^{-1}3d$ & 0.0066 & - & - & 0.0056 & 0.0045 & 0.0047 \\ 
$p_{3/2}^{-1}3d$ & 0.0130 & 0.0187 & 0.0199 & 0.0167 & 0.0131 & 0.0117 \\ 
$p_{1/2}^{-1}3d$ & 0.0069 & 0.0067 & 0.0069 & 0.0086 & 0.0064 & 0.0055 \\ 
$p_{3/2}^{-1}5s$ & 0.0068 & 0.0064 & 0.0066 & 0.0073 & 0.0060 & - \\ 
$p_{1/2}^{-1}5s$ & 0.0053 & 0.0043 & 0.0044 & 0.0050 & 0.0043 & - \\ \hline
\end{tabular}
\end{center}
\end{table}

\begin{figure}
%[tbp]
\centerline{\includegraphics*[scale=0.47]{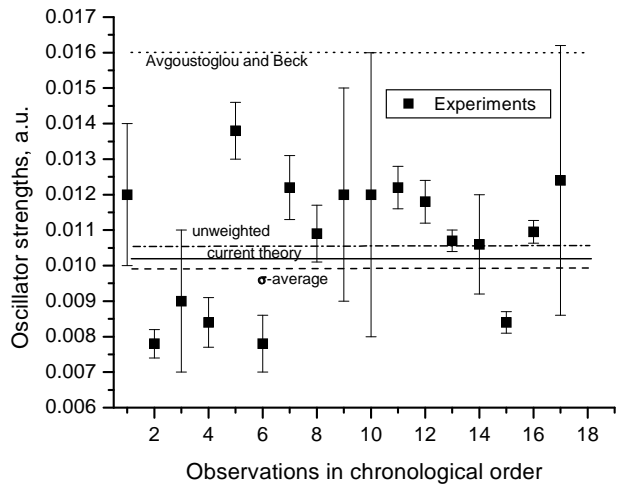}}
\caption{Comparison with experiment for oscillator strengths of the $[p_{3/2}^{-1}3s]_1$ state of neon}
\label{neonexp}
\end{figure}
\begin{table}[tbp]
\caption{References for experimental data shown in Fig.~\ref{neonexp}}
\label{graphtable}
\begin{center}
\begin{tabular}{llccc}
\hline
\hline
Obs. & Reference & Year & $f$ & $\sigma$ \\
\hline
1 & \citet{Kuhn}      & 1967&0.01200&0.00200 \\
2 & \citet{Lawrence}  & 1969&0.00780&0.00040 \\
3 & \citet{Geiger}    & 1970&0.00900&0.00200 \\
4 & \citet{Kernahan}  & 1971&0.00840&0.00070 \\
5 & \citet{Kazantsev} & 1971&0.01380&0.00080 \\
6 & \citet{Knystautas}& 1974&0.00780&0.00080 \\
7 & \citet{Bhaskar}   & 1976&0.01220&0.00090 \\
8 & \citet{Westerveld}& 1979&0.01090&0.00080 \\
9 & \citet{Aleksandrov}& 1983&0.01200&0.00300 \\
10 & \citet{Chornay}   & 1984 &0.01200&0.00400 \\
11 & \citet{Tsurubuchi}& 1990 & 0.01220 & 0.00060 \\
12 & \citet{Chan}      & 1992 & 0.01180 & 0.00060 \\
13 & \citet{Lightenberg}& 1994 & 0.01070 & 0.00030 \\
14 & \citet{suzuki}    & 1994 & 0.01060 & 0.00140 \\
15 & \citet{curtis}    & 1995 & 0.00840 & 0.00030 \\
16 & \citet{gibbson}   & 1995 & 0.01095 & 0.00032 \\
17 & \citet{neon:97}     & 1997 & 0.01240 & 0.00380 \\
\hline			     
\end{tabular}
\end{center}
\end{table}
Many experiments have disagreements in oscillator strengths far exceeding
the cited errors (see Fig.~\ref{neonexp} and Table~\ref{graphtable}): hence, 
for comparison, we give in Table~\ref{tab:neontr} two
statistical averages: the first is a weighted according to cited standard
deviations and the second is an unweighted average. For the 3s levels, the
experimental data compiled in Ref.~\cite{neon:98} and for the higher excited
levels in Ref.~\cite{neon:97} have been included in the averaging. Average
values obtained here are not necessarily the most accurate, but they serve
well for comparison and for a test of our probably less accurate calculated
values. 

A more careful analysis of experimental techniques to exclude systematic
errors, which are definitely present, is necessary; our values can provide
some guidance. For $p_{3/2}^{-1}3d$ states, since the energy separation of
the two states is small, experiments give the sum of the two oscillator
strengths, and the value 0.0196 rather than 0.0130 should be compared with the
experimental values 0.0187 (0.0199). In this table, we also compare our
theory with other semiempirical theories. Surprisingly, early calculations
by~\citet{neon:83} agree well with our calculations. A fair agreement,
considering the high sensitivity of these transitions to correlation
correction, is also obtained with the other theories in the table.

\section{Conclusions}

In this paper, we have introduced CI+MBPT theory for particle-hole states of closed-shell atoms.
A difficulty that the hole energy has
poor convergence is overcome with modifications of denominators in MBPT.
Good precision for hole states and for particle-hole states is illustrated
for many energy levels of neon. Apart from energies, our
theory is tested in calculations of oscillator strengths. 
Agreement with averaged experimental values is achieved.

\begin{acknowledgments}
The work of W. R. J. and I. M. S. was supported in part by National
Science Foundation Grant No.\ PHY-01-39928. 
\end{acknowledgments}

%\bibliography{disser}

\end{document}